\documentclass[a4paper,11pt]{article}
\pdfoutput=1 
\usepackage{amssymb,amsmath}

\usepackage{jcappub} 
\usepackage[T1]{fontenc} 
\usepackage{newtxtext,newtxmath}
\usepackage{ae,aecompl}
\usepackage{hyperref}
\usepackage{cleveref}
\usepackage{changes}

\usepackage{amssymb}	

\def\prd{Physical Review D}
\def\jcap{JCAP}

\def\mnras{MNRAS}

\def\sovast{Sov. Astron.}

\def\beq#1{\begin{equation}\label{#1}}
\def\eeq{\end{equation}}
\def\beqa#1{\begin{eqnarray}\label{#1}}
\def\eeqa{\end{eqnarray}}

\def\comment#1{\relax}

\def\pk#1{{\color{red}{ [\small{PK: #1}]}}}

\newcommand{\be}{\begin{eqnarray}}
\newcommand{\ee}{\end{eqnarray}}

\title{\boldmath Why the mean mass of primordial black hole distribution is close to 10$M
_\odot$}

\author[a,b]{A. Dolgov,}
\author[a,b,c]{and K. Postnov}

\affiliation[a]{ITEP,  Bol. Cheremushkinsaya ul., 25, 117218 Moscow, Russia}
\affiliation[b]{Novosibirsk State University,  Novosibirsk, 630090, Russia}
\affiliation[c]{Sternberg Astronomical Institute, Moscow M.V. Lomonosov State University,\\
 Universitetskij pr., 13,  Moscow 119234, Russia}

\emailAdd{dolgov@fe.infn.it}
\emailAdd{kpostnov@gmail.com}

\abstract{
It is shown that a mechanism of PBH formation \textbf{from high-baryon bubbles with log-normal mass spectrum} naturally leads to the central mass of the PBH  distribution close to ten solar masses independently of the model details. This result is in good agreement 
with observations.\\[2mm]
}

\begin{document}

\maketitle

\flushbottom

 \section{Introduction: PBHs from high-baryon bubbles}
 
The notion that primordial black holes (PBHs) with an extended mass spectrum can abundantly populate the 
contemporary universe becomes quite popular nowadays, since it has been strongly supported by observations (see, e.g., \cite{2016JCAP...11..036B,2020JCAP...01..031G}, and refs. \cite{2020arXiv200212778C,2018PhyU...61..115D} for a review). 
\textbf{The extended PBH mass spectrum} was suggested in 1993 in  paper~\cite{1993PhRvD..47.4244D}, where a {\bf new}
mechanism of massive and supermassive PBH 
production was proposed. According to this mechanism, the created PBHs have a log-normal mass spectrum:
\be
\frac{dN}{dM} = \mu^2 \exp \left[- \gamma \ln^2\left( \frac{M}{M_m} \right) \right]  ,
 \label{dN-dM}
\ee
where $\gamma$ is a dimensionless constant and the parameters $\mu$ and $M_m$ have dimension of mass or, what is the same, 
of inverse length (here the natural system of units with $c=k=\hbar =1$ is used). 

{\bf The conventional mechanism of PBH formation, originally proposed in refs~\cite{1967SvA....10..602Z, 1974MNRAS.168..399C}, assumes that at the radiation dominated  (RD) stage, the energy density fluctuations $\delta \rho /\rho$ at the cosmological horizon may accidentally reach unity. One can see that the piece of volume with such an
excessive density turns out to be inside its own gravitational radius, so it decouples from the Hubble flow  
and a black hole is created. Normally, it is believed that such PBHs have relatively low masses and 
rather sharp mass spectrum, even close to  delta-function.

In the model of ref.~\cite{1993PhRvD..47.4244D}, elaborated in more detail in the 
subsequent paper~\cite{2009NuPhB.807..229D}, the conditions for the PBH formation (but not PBHs themselves) could be created at the inflationary stage. A simplified discussion of the essential features of the scenario can be found in ref.~\cite{2018PhyU...61..115D}. The mechanism is based on the Affleck-Dine model of 
baryogenesis~\cite{1985NuPhB.249..361A}. According to this model, the baryon asymmetry is created by the 
decay of a classical complex scalar field $\chi$ with quartic potential with non-zero baryon number. This phenomenon is very similar to the appearance of the well known Higgs condensate in electroweak theory. 

Later, the field   $\chi$ decayed into quarks, probably with baryonic number conservation, leading to the 
baryon asymmetry in the sector of massless quarks. The natural value of the asymmetry, $\beta = n_B/n_\gamma$ (weere $n_B$ and $n_\gamma$ is the number density of baryons and CMB photons, respectively),
may be close to unity, much larger than the observed $\beta \approx 10^{-9}$.
This required searching for a physical mechanism
*preventing the creation of the classical field $\chi$ leading to a high baryon asymmetry
everywhere in the universe. 

The basic idea of ref.~\cite{1993PhRvD..47.4244D} is to introduce an interaction of the field $\chi$ with the inflaton field $\Phi$.
The effective mass squared, $m^2_\chi$, of 
the field $\chi$ is positive at the onset of inflation and during some time interval before inflation had terminated. However, 
when $\Phi$ is close to a certain value $\Phi_0$ the mass squared of $\chi$ became negative, exactly as in the electroweak theory. 
During the period when $m_\chi^2 < 0$, the former minimum of the potential at $\chi = 0$ becomes unstable and 
$\chi$ exponentially rises. The growth continues  
until $m^2_\chi$ changes the sign and $\chi$ starts to return to the minimum of the potential and to rotate in the two-dimensional complex plane of $\chi$. This rotation generates the baryonic number density inside the bubble which is related to 
the non-zero time-dependent phase $\theta$ of the field $\chi$ according to
\be
n_B^\chi = \dot \theta |\chi |^2.
\label{B-chi}
\ee

Since the gate to large values of $\chi$ are open only for a short time, the classical field 
$\chi$ reaches high values 
only in relatively small-size bubbles, where baryon
asymmetry could be close to unity, while the rest of the universe will have the observed low baryon asymmetry. 
Without significant fine tuning, the model parameters can be chosen such to ensure that the bubbles with 
high value of $\beta$ (and hence with large total baryonic number $B$) occupy a minor fraction of space.
We refer to such bubbles with high baryonic number as HBB. 

So we arrive at the
following picture. The bulk of the universe has the low baryon asymmetry $\beta \approx 10^{-9}$ with
relatively small but possibly astrophysically large bubbles with huge $B$. Since the carriers of
the baryonic number, quarks, in the very early universe were massless, the density contrast between HBBs and the bulk was
negligibly small. These are the so-called isocurvature fluctuations at very small scales.
The duration of inflation after the HBB formation enables creation of HBBs with a maximum 
mass of up to $(10^4 - 10^5) M_\odot$ \cite{2016JCAP...11..036B}.

Such a situation persisted until the QCD phase transition (QCD PT) when a condensate of the 
gluon field \cite{1979NuPhB.147..385S} 
with large negative vacuum-like energy, $\rho_{g} \approx -(300 {\rm MeV})^4$,  was formed.  The value of $\rho_g$ may differ roughly by  an order of
magnitude for different ways of calculations. As a result of the QCD PT,
quarks turned into massive protons and neutrons. 
The energy released due to gluon condensation was sufficient to  secure a large mass, $\sim 1$ GeV, 
to protons and neutrons.   At that stage, the density
of HBBs became larger than the background density, by the factor proportional to the magnitude of the
baryon asymmetry inside HBBs. For $\beta =1$ it would be exactly the mass inside the horizon at the QCD PT.
}

\section{PBH masses at the QCD phase transition}

The mechanism \cite{1993PhRvD..47.4244D,2009NuPhB.807..229D}
leads to a log-normal distribution of HBBs over their size at the moment of their creation,  
close to the end of inflaioin but still at the inflationary stage. 
If the mass of a HBB is larger than the mass inside the cosmological horizon at the QCD PT,  $M_{QCD}$,
PBHs with mass equal to $M_{QCD}$ or larger are produced with log-normal mass spectrum.
PBHs with smaller masses which might be possibly created by collapse of lighter HBBs would certainly
have different mass spectrum.

As described above, an essential feature of the model~\cite{1993PhRvD..47.4244D} is the creation of cosmologically small but astrophysically large bubbles with a 
very high baryonic density which would seed the PBH formation at the temperatures below $T_\mathrm{QCD}$.
Prior to the QCD phase transition, the density perturbations were vanishingly small because of zero quark masses, and
only isocurvature perturbations at 
small scales  were present. 

The temperature of the QCD PT from the free quark to the confined hadron phase is not well known. Is it thought to be 
$T_\mathrm{QCD} \approx 150 $ MeV for vanishing chemical potential of quarks, $\mu_q$, and can be 
somewhat 
 lower for $\mu_q \neq 0$, see e.g. the review~\cite{2008RvMP...80.1455A}\footnote{See also https://en.wikipedia.org/wiki/QCD\_matter}.

After the QCD PT, 
a collection of massless quarks ($m\ll T$) with the energy per particle equal to $3T$ is turned into massive
nucleons (protons and neutrons) with masses approximately 1 GeV, much larger than $T_\mathrm{QCD}$.
The matter becomes non-relativistic, so in about one Hubble time $\delta \rho / \rho$ in an HBB reaches unity, 
a BH is formed, and the bubble decouples from the Hubble flow.

Thus, starting from the QCD PT and later, the bubbles with masses equal to the mass inside the horizon would turn into PBHs. At the RD
cosmological stage, the mass inside the horizon is  $M_\mathrm{hor} = m_\mathrm{Pl}^2 t$, where
the Planck mass is 
$m_\mathrm{Pl} = 2.2 \times 10^{-5}$~g $= 2 \times 10^{43}$~s$^{-1}$ ={ $1.22\times 10^{19}$~GeV} and $t $
is the age of the universe in seconds.
Therefore,  $M_\mathrm{hor} = (4.4\times 10^{38}\,{\rm g})(t/[\mathrm{s}]) = (2.2 \times 10^5 M_\odot)(t/[\mathrm{s}])$. 

The age of the universe at the QCD PT can be  determined from the expressions of the cosmological energy density at the RD stage:
\be
\rho = \frac{3 m_\mathrm{Pl}^2}{32 \pi t^2} = \frac{\pi^2 g_*}{30} \,T^4,
\label{rho}
\ee
where $g_* \approx 40$ is the number of relativistic species in the primordial plasma above and close to $T_\mathrm{QCD}$.
Correspondingly, we find  
\be
t T^2 = 0.048 (40/g_*)^{1/2}\, m_\mathrm{Pl}\
\label{t-T2}
\ee
and hence $t /[\mathrm {s}] \approx 0.38 (T/\mathrm{MeV})^{-2} $. The mass inside the horizon at the QCD PT is thus
\be
M_\mathrm{hor}^{\mathrm{QCD}} \approx 8 M_\odot (100 \,{\rm MeV}/ T_\mathrm{QCD})^2.
\label{M-hor}
\ee

\section{Discussion and conclusion}

{\bf
The bubbles with radius equal to the cosmological horizon at QCD PT became black holes with the mass
equal to that inside the horizon. 
Larger BBHs, which entered under the horizon later, could create massive and supermassive BHs, with masses up to 
$(10^5-10^4) M_\odot$ depending upon the model details. These massive PBH could be the seeds for supermassive BHs in galaxies \cite{2016JCAP...11..036B} and intermediate-mass BHs in globular clusters \cite{2017JCAP...04..036D}.
}

The HBBs with masses smaller than $M_\mathrm{hor}^{\mathrm{QCD}}\approx m_\mathrm{Pl}^2 r_\mathrm{hor}/2$ 
would not produce black holes but rather
dense stellar-like objects 
with internal density close to that of neutron stars, $\rho \sim T_\mathrm{QCD}^4 \sim ({\rm 100\, \mathrm{MeV}})^4 \approx  3 \times 10^{13}\,  {\rm g/cm}^3 $
\cite{2015PhRvD..92b3516B}. Let us assume that their radius at the moment of QCD PT is a fraction $\varkappa <1 $ of the 
cosmological horizon, $r_\mathrm{HBB}= \varkappa r_\mathrm{hor}$.
Then their mass would be $M_\mathrm{HBB} =\varkappa^3 M_\mathrm{hor}^{\mathrm{QCD}} \sim \varkappa^3 r^3_\mathrm{hor}$. Correspondingly, their
gravitational radius would be smaller than their radius:
\be
r_g = 2M_\mathrm{HBB}/m^2_\mathrm{Pl}\sim \varkappa^3  2M_\mathrm{hor}^{\mathrm{QCD}}/m_\mathrm{Pl}^2  = \varkappa^3 r_\mathrm{hor} < r_\mathrm{HBB},
\label{r-g}
\ee
so they do not collapse into BH at the QCD PT. Further destiny of such smaller HBBs is interesting  but unclear \cite{2015PhRvD..92b3516B}. They may 
survive up to the present time as compact stars with unusual chemical composition enriched by metals because the big bang nucleosynthesis inside them proceeded with a very high baryon-to-photon ratio; they also may explode or collapse to BHs with
masses below $M_\mathrm{hor}^{\mathrm{QCD}}$ (\ref{M-hor}).

{\bf
In principle, such stellar-mass HBBs  could have some impact on the standard BBN in the bulk on the universe.
However, it seems to be
negligible, because their mass density is assumed to be smaller than the observed  density of
dark matter $\Omega_m\sim 0.3$. It means that their relative contribution to the energy density at QCD PT at $z\sim 10^{12}$ is not
larger than $10^{-8}$ and less than $10^{-6}$ at BBN. 
On the other hand, BBN inside HBBs would proceed quite differently due to a huge value of the baryon asymmetry $\beta $ (
which in the papers dedicated to BBN physics is 
usually denoted as $\eta$). According to calculations in refs.~\cite{2004PThPh.112..971M,2005PhRvD..72l3505M,2007PhRvD..75f8302M}, purely helium
HBBs or those enriched by heavy elements, possibly up to iron or even heavier, may originate. 
}

If the initial size of the bubbles with high baryonic number density was so large that
they reentered horizon after the QCD PT, then the PBH distribution in the contemporary universe would be log-normal with 
the  mid-mass $M_m$ higher or equal to $\sim 10 M_\odot$. We cannot exclude that $M_m$ might be much higher than  $10 M_\odot$,
which could happen if the original distribution over the bubble sizes has maximum at very large scales.
More natural
seems a log-normal distribution of the initial bubbles with the mid-size smaller than the horizon at the future QCD PT.   
In this case, we would expect to observe most of the PBHs with masses around  $10 M_\odot$, possibly some PBHs 
with smaller masses, and plenty of massive PBHs in the tail of the log-normal distribution (\ref{dN-dM}) with $M_m \sim 10 M _\odot$.

{\bf Remarkably, the chirp mass distribution of the coalescing black holes observed by LIGO/ Virgo interferometers 
can be well fitted by
a log-normal mass spectrum of the individual black holes with $M_m\sim 17 M_\odot$ \cite{2019JCAP...02..018R,2020arXiv200500892D}. Such a PBH mass spectrum is compatible with other observations as well \cite{2016JCAP...11..036B,2019arXiv190510972D}}.
In our opinion, it is a strong argument supporting the  PBH formation from high-baryon bubbles suggested in ref. \cite{1993PhRvD..47.4244D}. 
It is really striking that the value of $M_m$ is independent on the unknown high-energy details of the model \cite{1993PhRvD..47.4244D} and is determined by the low-energy/temperature cosmology. 

\section*{Acknowledgement}
\textbf{We thank the anonymous referees for 
careful reading of the paper and notes that helped us to improve the presentation of the paper}.
This work was supported by RSF grant  19-42-02004.

\providecommand{\href}[2]{#2}\begingroup\raggedright\endgroup

\end{document}